\documentclass[prb,preprint]{revtex4-1}

\usepackage{amsmath}
\usepackage{amssymb}
\usepackage{graphicx,color}
\usepackage{bbold,bm}
\usepackage{mathrsfs} 

\newcommand{\up}[1]{\left|\uparrow\right\rangle}
\newcommand{\down}[1]{\left|\downarrow\right\rangle}
\newcommand{\sgn}{\mathrm{sgn}}

\newcommand{\dbar}{\hbox{$d$\kern-0.6em\raise0.3em\hbox{$-$}}\hspace{-0.5mm}}

\providecommand{\ignore}[1]{}
\providecommand{\aucmnt}[1]{#1}
\newcommand{\be}{\begin{equation}}
\newcommand{\ee}{\end{equation}}
\renewcommand{\aucmnt}[1]{}

\newcommand{\Comment}[1]{}

\begin{document}

\title{On the role of self-adjointness in the continuum formulation of topological quantum phases}

\author{Mostafa Tanhayi Ahari}
\affiliation{Department of Physics, Indiana University, Bloomington,
IN 47405}

\author{Gerardo Ortiz}
\affiliation{Department of Physics, Indiana University, Bloomington,
IN 47405}

\author{Babak Seradjeh}
\email{babaks@indiana.edu}
\affiliation{Department of Physics, Indiana University, Bloomington,
IN 47405}

\begin{abstract}
Topological quantum phases of matter are characterized by an intimate
relationship between the Hamiltonian dynamics away from the edges and
the appearance of bound states localized at the edges of the system.
Elucidating this correspondence in the continuum formulation of
topological phases, even in the simplest case of a one-dimensional
system, touches upon fundamental concepts and methods in quantum
mechanics that are not commonly discussed in textbooks, in particular
the self-adjoint extensions of a Hermitian operator. We show how such
topological bound states can be derived in a prototypical
one-dimensional system. Along the way, we provide a pedagogical
exposition of the self-adjoint extension method as well as the role of
symmetries in correctly formulating the continuum, field-theory
description of topological matter with boundaries. Moreover, we show
that self-adjoint extensions can be characterized generally in terms of
a conserved local current associated with the self-adjoint operator.

\end{abstract}

\maketitle

\section{Introduction}

Phases of matter are operationally distinguished by their relevant 
physical observables. In quantum mechanics, these observables correspond
to linear, self-adjoint, and gauge-invariant operators
acting on a properly specified Hilbert space, whose elements represent the 
physical states of the system.~\cite{Jordan} In this formalism, 
measurable values of an observable are identified with the eigenvalues of
the corresponding operator, and since the result of a measurement is 
associated with a real number it is sensible to impose the condition of self-adjointness
to guarantee real-valued eigenvalues.  When the Hilbert state space is infinite-dimensional,
some operators may be unbounded and, therefore, need to be treated with special care; in
particular, determining the proper set of boundary conditions that
establishes their self-adjointness becomes a crucial step in their
definition.~\cite{BonFarVal01a} Specifically, for an unbounded operator to be self-adjoint, 
it is necessary but not sufficient for it to be Hermitian (symmetric). Therefore, a naive approach to
analyzing the eigenvalues and eigenvectors of Hermitian operators without properly
specifying the self-adjoint boundary conditions can lead to unexpected paradoxes,~\cite{BonFarVal01a}
resulting in wrong physical conclusions.

These considerations become quite important for the study of topological phases of
matter,~\cite{BH,She12a} where the properties of the physical system near boundaries are
closely linked with those of the bulk. In view
of the increased interest in topological phases, it is a goal of 
this work to illustrate the correct implementation of the theory of
self-adjoint extensions of operators to the continuum, or field-theory, description of
topological quantum matter. Fortunately, there is already a well-developed 
theory of self-adjoint extensions of Hermitian operators;~\cite{von,GitTuyVor12a}
however, its application to, and consequences for, the continuum description of topological
phases remain obscure. We believe this is in part due to the mathematically abstract nature 
of this theory. Thus, we also aim to provide a  formulation of the theory of
self-adjoint extensions that is more natural for physicists.

We will illustrate this method in the context of a
concrete and simple model of a topological phase. From a pedagogical
standpoint, topological phases provide a very natural setting for
learning and using self-adjoint extensions. It is a topic that brings
together several strands of modern physics and mathematics: many-body
quantum systems; symmetries of a physical system; topology; and
functional analysis. We present an accessible exposition
of these subjects in the context of a simple model of a ferromagnetic
insulator with spin-orbit interactions. We also clarify the role of
symmetry and boundary conditions in self-adjoint extensions of an
operator. In particular, we show that the notion of bulk-boundary
correspondence, i.e., a relationship 
between the topological properties
of the bulk phase and the character of its boundary states (see, e.g., 
Chap.~6 of Ref.~\onlinecite{BH}), depends on
the choice of self-adjoint extension. Finally, we provide a new and intuitive 
 physical interpretation of self-adjoint extensions in terms of a conserved
current associated with the self-adjoint operator.

The paper is organized as follows. In Sec.~\ref{sec:TM} we briefly
explain the notions of a topological state of matter and its concomitant
boundary states.  To illustrate these concepts, in Sec.~\ref{sec:QW}
we introduce a model of a one-dimensional lattice that exhibits normal
and topological phases over a range of parameters.  In
Sec.~\ref{sec:CL} we present a pedagogical derivation of the
continuum, field-theory description of this model with boundaries, and
point out the difficulty of dealing with unbounded operators. In
Sec.~\ref{sec:SAE} we review von Neumann's theory of self-adjoint
extensions of a Hermitian operator.  In Sec.~\ref{sec:TBS} we use this
theory to obtain the self-adjoint extensions of the continuum
model Hamiltonian of Sec.~\ref{sec:CL} and its topological bound states. We also discuss the effect
of a symmetry operation on the self-adjoint extension.  In
Sec.~\ref{sec:curr} we recast the self-adjoint extensions of
arbitrary Hermitian operators in terms of a conserved, spatially local
current.  We conclude in Sec.~\ref{sec:conc}. Technical details are
presented in four appendices.

\section{Topological Quantum Phases and Boundary States}\label{sec:TM}

The study of topological phases of quantum systems has flourished in
recent years, thanks to theoretical and experimental discoveries of
several families of such phases in artificial nanostructures as well as
bulk crystals.~\cite{BH,She12a} In contrast to the phases of matter
characterized by spontaneously broken symmetries, such as ferromagnets vs.\ paramagnets
or solids vs.\ liquids,\cite{Nishimori-Ortiz} topological phases of
electronic matter are characterized by the nontrivial topology of their
electrons' wave functions. In particular, a topological phase cannot be
distinguished from a normal phase by probing its bulk electronic
properties locally. For example, a topological insulator does not allow
the passage of electric current through it just like a normal insulator;
however, as one changes the momentum of the electrons, in a
topological insulator, their wave functions exhibit a nontrivial twist. This twist cannot be
undone by changing the parameters of the system unless the system is
brought to a critical phase at which it becomes a metal.~\cite{Note1}
This characterization also implies that at the boundary between a
topological phase and a normal phase, the electronic motion must undergo
a drastic change. For instance, the electronic states at the boundary between
a topological and a normal insulator are metallic. In other words, one
expects that at the boundaries of the system in a topological phase
there are bound states, protected by symmetry,  that emerge because of
the nontrivial topology of the bulk. This statement is commonly referred
to as the bulk-boundary correspondence.

Though crystals are described microscopically in terms of their lattice
structure, it is often more convenient to describe these systems in a
continuum, field-theory,  representation. Such a description is relevant for the
long-distance behavior of the system, which is what determines its quantum phases. 
It is also a more general
description since different lattice systems can end up having the same
continuum description and, thus, the same long-distance physics. As is
generically the case in quantum systems, continuum descriptions involve
unbounded operators such as the linear momentum. Usually this does not
present a major obstacle since one can regularize such operators by
restricting their action to square-integrable wave functions that vanish
sufficiently fast at infinity. However, in order to illustrate the
bulk-boundary correspondence in a topological phase, one must study the
properties of the system near the boundaries. In this case, one needs to
regularize the unbounded operators differently. Intuitively, this
regularization can be seen to involve the boundary conditions of the
electronic wave functions. Mathematically, it requires the notion of
self-adjoint extensions of unbounded Hermitian
operators.~\cite{Cap77a,BonFarVal01a,AraCouFer04a}

In order to illustrate these concepts, in the next section we introduce a simple 
lattice model in one spatial dimension, which exhibits a normal and a topological insulating phase. 
We derive the continuum Hamiltonian description of this model and study the spectral properties of the 
corresponding field theory. We shall see that correctly identifying the phases of the system 
in this continuum formulation requires the proper identification of boundary conditions related 
to the self-adjointness of the Hamiltonian operator.

\section{A topological quantum wire}\label{sec:QW}

Consider a system of  spin-$1/2$ fermions  moving along a line with a
periodic array of potential wells, for example electrons moving through
a chain of $N$ ions, subject to a magnetic field and
an internal spin-orbit interaction. When the separation between the ions $a$
is large enough so that only the quantum tunneling between nearest ions
is appreciable, one can model this system as a discrete chain with the
Hamiltonian,~\cite{She12a}
\begin{equation}\label{eq:Hlatt}
\hat{H}=\sum_{r=1}^{N}\left[
\frac\mu2\left(\hat{c}_{\uparrow r}^{\dagger}\hat{c}_{\uparrow r}^{\vphantom{\dagger}}-\hat{c}_{\downarrow r}^{\dagger}\hat{c}_{\downarrow r}^{\vphantom{\dagger}}\right)
+ w\left(\hat{c}^{\dagger}_{\uparrow r+1}\hat{c}_{\uparrow r}^{\vphantom{\dagger}} - \hat{c}^{\dagger}_{\downarrow r+1}\hat{c}_{\downarrow r}^{\vphantom{\dagger}}\right)
+ \frac\lambda2\left(\hat{c}_{\downarrow r+1}^{\dagger}\hat{c}_{\uparrow r}^{\vphantom{\dagger}} - \hat{c}_{\uparrow r+1}^{\dagger}\hat{c}_{\downarrow r}^{\vphantom{\dagger}}\right)
\right] + \text{h.c.}.
\end{equation}
Here, ``$+ \text{h.c.}$'' means one must add the Hermitian conjugate of
all the previous terms. We are using a second-quantized notation of the
creation and annihilation operators in the Fock space,
$\hat{c}_{sr}^\dagger$ and $\hat{c}_{sr}^{\vphantom{\dagger}}$, whose
action is to create and remove, respectively, an electron at ion site
$r\in\{1,2,\cdots\!, N\}$ with spin $s\in\{\uparrow,\downarrow\}$ along
the direction of the magnetic field. These operators obey the
anti-commutation relations
$\hat{c}_{sr}\hat{c}_{s'r'}+\hat{c}_{s'r'}\hat{c}_{sr}=0$ and
$\hat{c}_{sr}^{\vphantom{\dagger}}\hat{c}^{\dagger}_{s'r'}+\hat{c}^{\dagger}_{s'r'}\hat{c}_{sr}^{\vphantom{\dagger}}
= \delta_{ss'}\delta_{rr'}$, where $\delta$ is the Kronecker delta. The
operator product
$\hat{c}_{sr}^{\dagger}\hat{c}_{sr}^{\vphantom{\dagger}}$ counts how
many electrons (0 or 1) of spin $s$ are at site $r$. The operator
product $\hat{c}^{\dagger}_{s'r+1}\hat{c}_{sr}^{\vphantom{\dagger}}$
displaces an electron of spin $s$ from site $r$ to one of spin $s'$ at
site $r+1$. In this way, we can see that the parameter $\mu$ in the model
reflects the Zeeman energy splitting in the magnetic field; $w$
is an energy scale related to the strength of the spin-dependent quantum
tunneling between two nearest ions; and $\lambda$ is related to the
strength of the spin-orbit interaction. (Note the absence of a spin-independent
tunneling term, and the fact that the sum in Eq.~(\ref{eq:Hlatt}) does not include 
terms with $\hat{c}^{\vphantom{\dagger}}_{s N+1}$ and $\hat{c}^\dagger_{s N+1}$.) When the number of electrons is
$N$, the system may represent a ferromagnetic insulator with spin-orbit
coupling.  For simplicity, we shall take all the parameters to be real
and $w$ and $\lambda$ to be positive numbers. 

Note that this model has the following property: upon flipping the spin
 $\hat c_{\uparrow r}\mapsto \hat c_{\downarrow r}$ and $\hat
 c_{\downarrow r}\mapsto \hat c_{\uparrow r}$, one finds $\hat H\mapsto
 - \hat H$. Denoting the  operation with the unitary map $\hat S=(\hat
 S^\dagger)^{-1}$, we can express this as $\hat S\hat H \hat S^\dagger =
 -\hat H$. This means that if we flip the spins in any energy eigenstate
 $|E\rangle$ with energy $E$, $\hat H|E\rangle = E|E\rangle$, we obtain
 an eigenstate with energy $-E$, since
\begin{equation}
\hat H (\hat S|E\rangle) = -(\hat S\hat H \hat S^\dagger) \hat S|E\rangle = -\hat S\hat H |E\rangle = -E(\hat S|E\rangle).
\end{equation}
Therefore the spectrum of $\hat H$ is symmetric around $E=0$.

\begin{figure}[t]
\begin{center}
\includegraphics[width=3.5in]{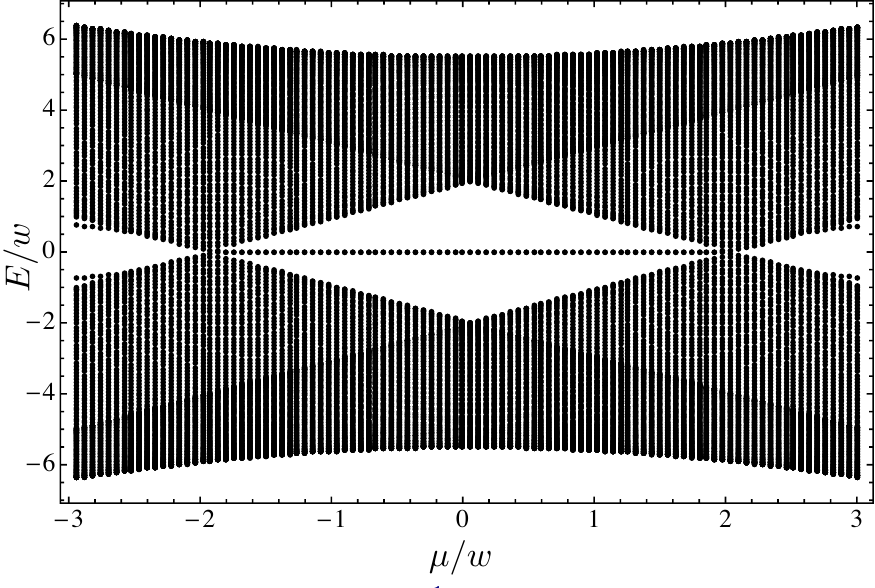}
\caption{The single-particle energy spectrum of the quantum wire shows the quantum phases of the system. Here, $N=150$ and $\lambda/w=5.5$. In the topological insulator phase, $|\mu/2w|<1$, a pair of zero-energy bound states is found localized at the two edges of the lattice. They spilt away in the normal insulator phase.}\label{fig:phase}
\end{center}
\end{figure}

This quantum wire is a prototypical system that displays different
quantum phases upon changes in the parameters of the Hamiltonian that
are distinguished by their topological properties. This can be seen more
easily if we rotate the spin basis to a direction orthogonal to the
magnetic field. In the new basis, the annihilation operators are
\begin{align}
\hat{b}_{\uparrow r} =& \frac{\hat{c}_{\uparrow r}+\hat{c}_{\downarrow r}}{\sqrt2},\\
\hat{b}_{\downarrow r} =& \frac{\hat{c}_{\uparrow r}-\hat{c}_{\downarrow r}}{\sqrt2}.
\end{align}
After some straightforward algebra, we obtain the Hamiltonian in the new basis as
\begin{equation}
\hat{H}=\sum_{r=1}^{N}\left[
\mu \hat{b}_{\uparrow r}^\dagger \hat{b}_{\downarrow r}^{\vphantom{\dagger}} 
+ \left(w + \frac\lambda2\right) \hat{b}_{\uparrow r+1}^\dagger \hat{b}_{\downarrow r}^{\vphantom{\dagger}}
+ \left(w - \frac\lambda2\right) \hat{b}_{\downarrow r+1}^\dagger \hat{b}_{\uparrow r}^{\vphantom{\dagger}}
\right] + \text{h.c.}
\end{equation}
Now we can easily see that when the parameters are at the ``sweet spot'' $\mu=0$ and $w=\lambda/2$, the Hamiltonian reduces to $\hat H = \lambda\sum_{r=1}^{N-1} \hat{b}_{\uparrow r+1}^\dagger \hat{b}_{\downarrow r}^{\vphantom{\dagger}} +\text{h.c.}$  Therefore, the two operators $\hat b_{\uparrow 1}^\dagger$ and $\hat b_{\downarrow N}^\dagger$ (and their Hermitian conjugates) do not appear in the Hamiltonian and $[\hat H, \hat b_{\uparrow 1}]=[\hat H,\hat b_{\downarrow N}]=0$. This means that the bound states created by $\hat b_{\downarrow N}^\dagger$ and $\hat b_{\uparrow 1}^\dagger$ at the two ends of the wire are degenerate eigenstates of the Hamiltonian (with zero energy).

One can show that even away from the sweet spot, there are two bound eigenstates of the Hamiltonian as long as $|\mu|<2w$. The energy of these bound states is exponentially small in the system size, $E_b\sim e^{-Na/\xi}$. Here $a$ is the lattice spacing between nearest-neighbor ions and $\xi$ is a length scale that depends on system parameters. In the thermodynamic limit, $N\to\infty$, we recover two asymptotically exact zero-energy bound states. The existence or absence of these bound states marks the topological or trivial phases of the system. The existence of these phases is illustrated in Fig.~\ref{fig:phase}, which shows the single-particle energy spectrum of the lattice Hamiltonian as $\mu$ is varied.

Instead of studying the phases on the lattice, we focus on the thermodynamic limit in the following and derive a continuum Hamiltonian. This approach has the advantage of simplifying the solutions since it is often easier to solve continuous differential equations rather than discrete difference equations. Also, many different lattice models can have the same continuum description in the thermodynamic limit. Thus, by keeping only the relevant long-distance, universal features of the system in the continuum limit, we will gain a more general perspective on the potential phases of the system.

\section{The continuum limit}~\label{sec:CL}

We shall take the continuum limit of the lattice model as the lattice constant $a\rightarrow0$ while the length $L=Na$ is fixed.  We leave open the choice of whether the length is finite, semi-infinite, or infinite.  In this limit, the discrete site index $r$ is turned into a continuous position $x$. A simple way to deduce the relation between fermion operators $\hat c_{sr}$ in the lattice and the field operator $\hat \Psi_s(x)$ in the continuum is to demand the proper anti-commutation relations
\begin{subequations}\label{eq:commCL}
\begin{align}
\hat{\Psi}_s(x)\hat{\Psi}_{s'}(x')+\hat{\Psi}_{s'}(x')\hat{\Psi}_s(x) &=0, \\
\hat{\Psi}_s^{\vphantom{\dagger}}(x)\hat{\Psi}_{s'}^{\dagger}(x')+\hat{\Psi}_{s'}^{\dagger}(x')\hat{\Psi}_s^{\vphantom{\dagger}}(x) &= \delta_{ss'}\delta(x-x'),
\end{align}
\end{subequations}
where $\delta{(x-x')}$ is the Dirac delta function.~\cite{Jordan} Thus, the fermion field $\hat\Psi_s(x)$ must have the dimension of (length)$^{-1/2}$. Defining $x=ar$, $x'=ar'$, and $\lim_{a\to0}(\delta_{rr'}/a) = \delta(x-x')$, we see that the proper definition of continuum operators has the form $\hat\Psi_s(x) = \lim_{a\to0}(\hat{c}_{sr}/\sqrt{a})$. In the limit $a\to0$ we can expand 
\begin{equation}
\frac1{\sqrt a}\hat c_{sr+1}\to\hat \Psi_s(x+a) = \hat \Psi_s(x) + a\frac{d}{d x}\hat\Psi_s(x)+O(a^2). 
\end{equation}
Also in this limit $a\sum_r \to \int dx$. Replacing these limits in the Hamiltonian, Eq.~(\ref{eq:Hlatt}), we find several terms that, to first order in $a$, vanish upon integration. In particular, terms of the form 
\begin{equation}
a w \int \hat\Psi_s^\dagger(x)\frac{d}{dx}\hat\Psi_s^{\vphantom{\dagger}}(x)\,dx+\text{h.c.}= 
a w \int \frac{d}{dx} \left[ \hat\Psi_s^\dagger(x) \hat\Psi_s^{\vphantom{\dagger}}(x) \right] d x 
\end{equation}
vanish in the limit $a\to0$ for finite $w$.
After some straightforward algebra, we find the continuum Hamiltonian
\begin{equation}
\hat H\to \hat H_c = \int\hat{\Psi}^\dagger(x)\mathcal{H}\hat\Psi(x)\,dx,
\end{equation}
where
\begin{equation}
\hat{\Psi}(x) = \begin{pmatrix} \hat\Psi_\uparrow(x) \\ \hat\Psi_\downarrow(x) \end{pmatrix}, \quad
\mathcal{H} = \begin{pmatrix} m & v \frac{d}{dx} \cr - v \frac{d}{dx} & -m \end{pmatrix},
\end{equation}
with $m=2w+\mu$ and $v=a\lambda$. Note that in this limit we keep $v$ finite. 
Using the Pauli matrices 
\begin{equation}
\sigma_x=\biggl(\begin{array}{cc} 0 & 1 \\ 1 & 0 \end{array}\biggr), \quad \sigma_{y}=\biggl(\begin{array}{cc} 0 & -i \\ i & 0 \end{array}\biggr) , \quad \sigma_{z}=\biggl(\begin{array}{cc} 1 & 0 \\ 0 & -1 \end{array}\biggr), 
\end{equation}
the matrix $\mathcal{H}$ could be written in a compact form as
\begin{equation}\label{eq:H}
\mathcal{H} =iv\sigma_y\frac{d}{dx}+m\sigma_z.
\end{equation}
We note that this is the Hamiltonian leading to the one-dimensional Dirac equation.~\cite{Coutinho88}
The spin-flip operation of the lattice Hamiltonian, $\hat{S}$ of Sec.~\ref{sec:QW}, is now implemented by the Pauli matrix $\mathcal{S}=\sigma_x$ and $\mathcal{SHS^\dagger=-H}$. 

The matrix $\mathcal{H}$ is an operator in the position basis. Its eigenvalues $E$ and eigenstates~\cite{Transpose} 
$\phi_E(x)=( \phi_{\uparrow E}(x) \ \phi_{\downarrow E}(x) )^{\sf T}$
are found by solving the linear system of differential equations
\begin{equation}\label{eq:1stH}
\mathcal{H}\phi_E(x) = E\phi_E(x).
\end{equation}
Using these solutions, we can define the energy creation operators
\begin{equation}\label{eq:psiE}
\hat\Psi_E^\dagger = \int \left[ \phi_{\uparrow E}(x) \hat{\Psi}_{\uparrow}^\dagger(x) + \phi_{\downarrow E}(x)  \hat{\Psi}_{\downarrow}^\dagger(x) \right] dx.
\end{equation}
As we show in Appendix~\ref{app:ladderCL}, the $\hat\Psi_E^\dagger$ are ``ladder operators'' that satisfy the commutation relations $[\hat H_c,\hat\Psi_E^\dagger] = E\hat\Psi_E^\dagger$. In this way, the full spectrum of the continuum Hamiltonian can be obtained by solving the first-quantized eigenvalue equation~(\ref{eq:1stH}). We will concentrate on studying the properties of $\mathcal{H}$ in the rest of this paper.

The careful reader may have noticed that the energy $E$ in Eq.~(\ref{eq:1stH}) 
can be unboundedly large. For instance, consider a solution such as
\begin{equation}
\phi_{E}(x)=e^{ikx}\Biggl(\begin{matrix} 1 \\ \frac{E-m}{i v k} \end{matrix}\Biggr),
\end{equation}
with eigenvalue $E=\pm\sqrt{v^2 k^2+m^2}$.  The energy $E$ can be an 
indefinitely large number for large $k$.  Thus $\mathcal{H}$ and the continuum Hamiltonian $\hat H_c$, unlike the lattice Hamiltonian, are unbounded operators. More surprisingly, even though the Hamiltonian is formally Hermitian, in a finite geometry ($x<\infty$ and/or $x>-\infty$) the energy eigenvalues can be imaginary numbers. We can see this by replacing $k$ with $i\kappa$ in the above solution. Such a solution would be acceptable since $e^{-\kappa x}$ need not diverge in a finite geometry for a proper choice of $\sgn(\kappa)$. When $|v\kappa|>|m|$, the energy is a purely imaginary number. 

The presence of imaginary eigenvalues is problematic for the correspondence of the Hamiltonian to observable energy. Formally, it signals the loss of self-adjointness of the Hamiltonian operator. To restore self-adjointness, the Hamiltonian needs to be properly defined on a subset of the Hilbert space such that no imaginary eigenvalues exist in the spectrum. In the following two sections we will introduce the mathematical technique to deal with this problem for a general unbounded operator, and find the appropriate subset that resolves this issue for our continuum Hamiltonian.

The following exercise shows that there is a second continuum field operator that one can define as $\hat{\Phi}_s(x) = \lim_{a\to0}(-1)^r\hat{c}_{sr}/\sqrt{a}$.
\begin{quote}
\textbf{Exercise 1.} By expanding the field operators $\hat{\Phi}_s$ in the limit $a\to0$, show that the continuum Hamiltonian for 
$\hat\Phi=( \hat\Phi_\uparrow \ \hat\Phi_\downarrow)^{\sf T}$ 
is 
\begin{equation}
\hat H \to \int\hat{\Phi}^\dagger(x)\left[ -iv\sigma_y\frac{d}{dx}+(\mu-2w)\sigma_z \right]\hat\Phi(x)\,dx.
\end{equation}
\end{quote}
Thus, as the analysis in the following sections shows, the continuum limit in terms of $\hat\Phi$ becomes important for the topological phases of the system when $\mu-2w$ changes sign. For simplicity, we shall assume $2w>\mu$ in the following.

\section{The self-adjoint extensions of a Hermitian operator}\label{sec:SAE}

The state of a quantum system is specified by a vector $\phi$ in a Hilbert 
space $\mathscr{H}$.~\cite{Jordan} The Hilbert space is equipped with an inner product $\langle\psi|\phi\rangle$ that is a complex-valued, positive-definite, sesquilinear function on $\mathscr{H}\times\mathscr{H}$ that maps any two states $\phi$, $\psi \in \mathscr{H}$ to a complex number $\langle \psi|\phi \rangle=\langle \phi|\psi \rangle^*$. The inner product defines a norm for the Hilbert space, $\|\phi\|\equiv\sqrt{\langle \phi|\phi \rangle}$. For example, given an interval $[L_1,L_2]\subset\mathbb{R}$, the Hilbert space $\mathscr{H}=\mathscr{L}^2[L_1,L_2]$ is the set of all square-integrable functions with the inner-product $\langle\psi|\phi\rangle=\int_{L_1}^{L_2} \psi^*(x)\phi(x)\,dx$, i.e., those functions with $\|\phi\|<\infty$. 

Physical observables of a quantum system are specified by linear self-adjoint operators.~\cite{Jordan} For an operator $A$ to be self-adjoint, it primarily must be Hermitian (or symmetric). An operator $A$ is defined to be Hermitian if 
\begin{equation}\label{eq:Herm}
\langle\psi| A\phi\rangle =\langle A\psi|\phi\rangle
\end{equation}
for all vectors $\psi$ and~$\phi$ in its domain.
In many textbooks on quantum mechanics, no distinction is made between Hermitian and self-adjoint operators. Indeed, one need not make such a distinction for bounded operators. (A bounded operator $B$ is one for which there exists a positive number $M$ such that $\|B\phi\|<M\|\phi\|$ for all vectors~$\phi$ in the Hilbert space.) However, the distinction between Hermitian and self-adjoint operators becomes crucial for unbounded operators. This is so because for unbounded operators the designation of the domain of the operator $A$, $\mathscr{D}(A)$,  which is  a subspace of the full Hilbert space $\mathscr{H}$, $\mathscr{D}(A) \subset \mathscr{H}$, generally becomes an important part of its definition.

For Hermitian operators, the connection between unboundedness and a restricted domain is quite general. In fact, according to the Hellinger-Toeplitz theorem,~\cite{Reed-Simon1} a Hermitian operator $B$ with the domain $\mathscr{D}(B)=\mathscr{H}$ is bounded. It is a direct corollary of this theorem that an unbounded and Hermitian operator cannot be defined on the whole Hilbert space. Let us illustrate this point by an example. Consider the position operator, $X$, on the Hilbert space $\mathscr{L}^2[0,L]$ defined as $(X\phi)(x)=x\phi(x)$ for $\phi \in \mathscr{L}^2[0,L]$. Equation~(\ref{eq:Herm}) is satisfied since
\begin{equation}\label{eq:example}
\langle \psi | X\phi \rangle = \int_0^L\psi^*(x) x \phi(x)\,dx= \int_0^L (x \psi(x))^* \phi(x)\,dx = \langle X\psi | \phi \rangle.
\end{equation}
Thus, the position operator is a Hermitian operator. In this case, $X$ is a bounded operator since for any $\phi\in\mathscr{L}^2[0,L]$, $\|X\phi\| < M\|\phi\|$ with $M=L$. Now, if instead we consider the whole real line, $\mathbb{R}$, the position operator defined on the Hilbert space $\mathscr{L}^2(\mathbb{R})$ continues to be Hermitian, but is no longer bounded.
Now, consider $\phi\in\mathscr{L}^2(\mathbb{R})$ such that 
$\phi(x)=x/(x^2+1)$. It is clear that 
$(X\phi)(x)=x^2/(x^2+1)$ is not square-integrable and therefore $X\phi$ is ill-defined. Therefore, in this case $X$ cannot be defined on the whole Hilbert
space and one must find an appropriate domain for $X$, such as $\mathscr{D}(X) = \{\phi\in \mathscr{L}^2(\mathbb{R}): \|X\phi\|<\infty \}$. This means $\phi(x)\to0$ for $x\to\infty$ faster than $x^{-3/2}$. As we shall discuss in detail below, restrictions on the domain of an operator are commonly expressed in terms of boundary conditions on the state vectors.

In order to define a self-adjoint operator properly, we first need to define the adjoint of an operator and its domain. For an operator $A:\mathscr{D}(A)\to\mathscr{H}$ and $\phi\in\mathscr{D}(A)$, the adjoint $A^\dagger$ and its domain of states $\psi\in\mathscr{D}(A^\dagger)$ are defined by
\begin{equation}\label{eq:adj}
\langle \phi | A^\dagger \psi\rangle := \langle A\phi | \psi \rangle.
\end{equation}
(This defines $A^\dagger\psi$ uniquely only if the operator $A$ is ``densely defined;'' see Appendix~\ref{app:dense} for a discussion.) A self-adjoint operator   $A$ is then defined as one that is Hermitian \emph{and} has the same domain as its adjoint,  $\mathscr{D}(A^\dagger)=\mathscr{D}(A)$.

As an example, consider the momentum operator, $P$, on the Hilbert space $\mathscr{L}^2[0,L]$, defined as $(P\phi)(x) = -id\phi/dx$ (we use units such that Planck's constant $\hbar=1$). Since the derivative of a square-integrable function can be arbitrarily large, $P$ is
unbounded. For its domain we take $\mathscr{D}(P)=\{\phi\in\mathscr{L}^2[0,L] : \phi(x) \text{ is absolutely continuous and} \phi(0)=\phi(L)=0\}$.\cite{Note1-5}  (One can show that with this definition, $P$ is densely defined; see Appendix~\ref{app:dense}.) Simple algebra shows that Eq.~(\ref{eq:adj}) holds true if we have 
\begin{subequations}
\begin{align}
(P^{\dagger}\psi)(x) &= -i\frac{d}{d x}\psi(x),\\
\psi^*(L)\phi(L) &= \psi^*(0)\phi(0).
\end{align}
\end{subequations}
The first equation shows that the momentum operator is a Hermitian operator. 
However, the second equation shows that $\psi(0)$ and $\psi(L)$ could be nonzero and still Eq.~(\ref{eq:adj}) would be satisfied. That is, $\mathscr{D}(P^\dagger) = \{ \psi\in\mathscr{L}^2[0,L] : \psi(x) \text{ is absolutely continuous} \} \supset \mathscr{D}(P)$. Therefore, $P$ is not self-adjoint. Some puzzling consequences of the lack of self-adjointness for $P$ are discussed in Ref.~\onlinecite{GitTuyVor12a}. An important consequence for our purposes is that the domain of the adjoint operator $P^\dagger$ is so unconstrained that it admits \emph{imaginary} eigenvalues. That is, one can find states $\psi_\pm\in\mathscr{D}(P^\dagger)$ such that $P^\dagger\psi_\pm = \pm i\eta\psi_\pm$ for $\eta>0$. Indeed, one can easily show that the two states
\begin{equation}\label{eq:Ppsi}
\psi_+(x) = e^{-\eta x}, \quad
\psi_-(x) = e^{-\eta(L-x)},
\end{equation}
are such eigenstates with equal norm, $\|\psi_+\|=\|\psi_-\|$.~\cite{Note2}

As seen in this example, the domain of the Hermitian operator $A$ is no larger than the domain of its adjoint, $A^\dagger$. This means that in order to make the operator $A$ self-adjoint one must \emph{extend} its domain in a way that shrinks the domain of the adjoint to match it, and eliminate the imaginary eigenvalues from the spectrum of the adjoint operator. If the procedure is successful, we find a self-adjoint extension of the Hermitian operator $A$.  This procedure was made systematic by von Neumann by a method known as the deficiency indices method.

In this method, given a Hermitian, densely defined, and closed operator $A:\mathscr{D}(A)\to\mathscr{H}$ (see Appendix~\ref{app:dense} for the definitions), one first finds the adjoint operator $A^\dagger$ and its domain $\mathscr{D}(A^\dagger)$ and then solves for the imaginary eigenstates of the adjoint operator. These solutions define the two \textit{deficiency} subspaces $\mathscr{K}_{\pm}=\{\psi_{\pm} \in \mathscr{D}(A^{\dagger}), A^{\dagger} \psi_{\pm}= \pm i\eta\psi _{\pm} \}$. The dimensions of these subspaces, $n_\pm=\dim(\mathscr{K}_\pm)$, are the deficiency indices of~$A$.

\begin{quote}
\textbf{Deficiency indices theorem} (Von Neumann).\cite{Reed-Simon1,Reed-Simon2,von} For a Hermitian, densely defined, and closed operator $A$,
\begin{enumerate}
\item If $n_+ = n_- =0$, $A$ is self-adjoint.
\item If $n_+\neq n_-$, $A$ is not self-adjoint and it has no self-adjoint extensions. 
\item If $n_+=n_-=n>0$, $A$ has infinitely many self-adjoint extensions $A_{\mathcal{U}}$ parametrized by $n\times n$ unitary matrix maps $\mathcal{U}:\mathscr{K}_+ \to \mathscr{K}_-$. For a choice of  $\,\mathcal{U}$, $\mathscr{D}(A)\subset\mathscr{D}(A_\mathcal{U})\subset\mathscr{D}(A^\dagger)$,
\begin{align}\label{eq:domU}
\mathscr{D}(A_{\mathcal{U}}) &= \{\psi=\phi+\psi_++\mathcal{U}\psi_+: \phi \in \mathscr{D}(A), \psi_+ \in \mathscr{K}_+\}, \\
A_{\mathcal{U}} \psi &= A^\dagger \psi = A\phi + i\eta\psi_+ - i \eta\: \mathcal{U}\psi_+.
\end{align}
\end{enumerate}
\end{quote}
Note that even though $\eta$ appears explicitly in the deficiency subspaces, the family of self-adjoint extensions is independent of the parameter $\eta$.~\cite{GitTuyVor12a} So, sometimes the choice $\eta=1$ is made. However, we will keep $\eta$ for dimensional purposes.

Considering the example of the momentum operator $P$ on $\mathscr{L}^2[0,L]$, we see that $n_+=n_-=1$. Therefore there is a one-parameter family of self-adjoint extensions $P_{\mathcal{U}}\equiv P_\theta$ with $\mathcal{U}:\psi_+\mapsto e^{i\theta}\psi_-$ and $\psi_\pm$ given in Eq.~(\ref{eq:Ppsi}), such that
\begin{align}
P_\theta\psi &= -i\frac{d}{dx}\psi, \\\mathscr{D}(P_\theta) &= \{ \psi=\phi + c(\psi_+ + e^{i\theta}\psi_-) : \phi(0)=\phi(L)=0, 
c\in\mathbb{C} \}. \label{eq:Ptheta}
\end{align}
Note that $\psi\in\mathscr{D}(P_\theta)$ satisfies a new boundary condition, $\psi(L)=e^{i\beta}\psi(0)$, with $e^{i\beta} = (e^{-\eta L} + e^{i\theta})/(1+e^{-\eta L} e^{i\theta})$.

While this method is systematic, it is rather abstract. In Sec.~\ref{sec:curr} we will 
introduce a more intuitive way to find the proper domain of self-adjoint extensions in terms of boundary conditions and the local current associated with the corresponding self-adjoint operator.

\section{Self-adjoint extensions and topological bound states}\label{sec:TBS}

We will now apply von Neumann's deficiency indices method to obtain 
self-adjoint extensions of the continuum Hamiltonian $\mathcal{H}$, Eq.~(\ref{eq:H}), in 
semi-infinite and finite-size wire geometries. In each case, 
we will show that the appropriately extended domains admit 
additional bound states in the topological phase, and we will relate their 
properties to the boundary conditions at the ends of the wire. We will also see that demanding that the spectral symmetry, given by the spin-flip operation ${\cal S}$ of Sec.~\ref{sec:CL}, be preserved by the self-adjoint extension forces the bound states to have zero energy.

\subsection{Semi-infinite wire}

The simplest geometry with a boundary is the semi-infinite wire, $x\in[0,\infty)$, 
with just one physical boundary. In order to apply the deficiency theorem, we choose the domain
\begin{equation}
\mathscr{D}(\mathcal{H})=\left\{\phi=\Biggl(\begin{matrix}\phi_\uparrow \\ \phi_\downarrow \end{matrix}\Biggr):\phi_s(x)\in\mathscr{L}^2[0,\infty), \ \phi_s(0)=0\right\} ,
\end{equation}
where $s$ is the spin. (This domain is dense; the proof is similar to the example given in Appendix~\ref{app:dense}). With this choice, the adjoint operator has the domain
$\mathscr{D}(\mathcal{H}^\dagger)=\{\psi: \psi_s(x)\in\mathscr{L}^2[0,\infty)\}$.
Now we can find all possible extensions $\mathcal{H_U}$ such that $\mathscr{D}(\mathcal{H_U})=\mathscr{D}(\mathcal{H^\dagger_U})$. 

Next, we solve for $\mathcal{H}^{\dagger}\psi_{\pm}=\pm i\eta \psi_{\pm}$. It is easy to see that
\begin{equation}
\psi_\pm(x) = e^{-|\epsilon|x/v} \Biggl( \begin{array}{c} 1 \\ e^{\pm i\delta} \end{array} \Biggr),
\end{equation}
where $ |\epsilon| e^{i\delta}=m-i\eta$. (Notice that $\delta\neq0,\pi$ and $\|\psi_+\|=\|\psi_-\|$.)
So the deficiency indices are $n_+=n_-=n=1$. The domain 
of the extended operator $\mathcal{H_U}\equiv \mathcal{H}_\theta$ is parametrized by the one-parameter mapping $\mathcal{U}:\psi_+ \mapsto e^{i\theta}\psi_-$ as
\begin{equation}
\mathscr{D}(\mathcal{H}_\theta)=\{\psi=\phi+c(\psi_++e^{i \theta}\psi_-) : \phi(0)=0,\ c\in\mathbb{C} \}.
\end{equation}

The domain $\mathscr{D}(\mathcal{H}_\theta)$ is equivalently characterized by a new boundary condition. To see this, note that for an arbitrary state $\psi\in\mathscr{D}(\mathcal{H}_\theta)$ we have
\begin{equation}
\psi(0)=\Biggl(\begin{matrix} \psi_\uparrow(0) \\ \psi_\downarrow(0) \end{matrix}\Biggr) = c\Biggl(\begin{matrix} 1+e^{i\theta} \\ e^{i\delta}+e^{i\theta}e^{-i\delta} \end{matrix}\Biggr).
\end{equation}
Thus
\begin{subequations}\label{eq:BCsemi}
\begin{align}
\psi_\uparrow(0) = \Lambda \psi_\downarrow(0&),\\
\Lambda = \frac{\cos(\theta/2)}{\cos(\theta/2-\delta)} &\in\mathbb{R}.
\end{align}
\end{subequations}

Since we have extended the domain of $\mathcal{H}$, we may expect the spectrum of $\mathcal{H}_\theta$ to depend on the choice of $\theta$ or, equivalently, $\Lambda$. Indeed, one can easily check that for $m\Lambda>0$ there is always a bound state,
\begin{equation}
\psi_b(x) = \exp\Bigl(-\frac{m}{v}\frac{2\Lambda}{\Lambda^2+1}x\Bigr)\Biggl(\begin{matrix} \Lambda \\ 1 \end{matrix}\Biggr),
\end{equation}
with energy eigenvalue
\begin{equation}
E_b=\frac{\Lambda^2-1}{\Lambda^2+1}m.
\end{equation}
Note that $-|m|<E_b<|m|$, i.e., it is within the energy gap of the bulk states.

As the careful reader may infer from the dependence of the spectrum on
the choice of self-adjoint 
extension, other physical properties may also depend on this 
choice. An important case is the 
effect of a symmetry operation. For example, what happens to the spectral symmetry given by the spin-flip operation $\mathcal{S}=\sigma_x$ of Sec.~\ref{sec:CL}? It is easy to see that under the operation $\mathcal{S}$ the boundary condition characterizing $\mathscr{D}(\mathcal{H}_\theta)$ is mapped as  
$\Lambda\mapsto \Lambda'=1/\Lambda$. Therefore, $\mathcal{S}$ is not in general a symmetry of 
the spectrum and the spectrum may not be symmetric around zero. Indeed, the existence of a single bound state with $E_b\neq0$ is possible only if the spectrum is not symmetric. The spectral
symmetry is restored for $\Lambda'=\Lambda=\pm1$; for $\Lambda=\sgn(m)$ the spectrum has a zero-energy bound state, $E_b=0$. Therefore, whether or not a symmetry of the bulk lattice is a symmetry of the continuum description depends on the choice of self-adjoint extension.

\subsection{Finite wire}

We now consider a more realistic wire geometry:  a wire with a finite
length~$L$. We choose the domain of the Hamiltonian in Eq.~(\ref{eq:H}) to be 
\begin{equation}
\mathscr{D}(\mathcal{H})=\{\phi:\phi_s\in\mathscr{L}^2[0,L],\ \phi_s(0)=\phi_s(L)=0\},
\end{equation}
where $s$ is the spin. (One can show that this is a dense domain similar to previous cases.) With this choice, the domain of the adjoint operator is
\begin{equation}
\mathscr{D}(\mathcal{H}^\dagger)=\{\psi: \psi_s\in\mathscr{L}^2[0,L]\}.
\end{equation}
The deficiency indices are determined by solving the imaginary eigenvalue equation $\mathcal{H}^{\dagger}\psi_{\pm}=\pm i\eta \psi_{\pm}$.  We now find \textit{two} independent solutions for each sign,
\begin{equation}
\psi^{(1)}_\pm(x) = \, e^{-|\epsilon|x/v} \Biggl(\begin{matrix} 1 \\
e^{\pm i\delta} \end{matrix}\Biggr), \quad
\psi^{(2)}_\pm(x) = e^{-|\epsilon|(L-x)/v} \Biggl(\begin{matrix} 1 \\
- e^{\pm i\delta} \end{matrix}\Biggr),
\end{equation}
with $|\epsilon|e^{i\delta}=m-i\eta$ as before. (Note also that $\|\psi_+^{(1)}\|=\|\psi_-^{(1)}\|=\|\psi_+^{(2)}\|=\|\psi_-^{(2)}\|$.) The deficiency spaces $\mathscr{K}_\pm=\{ \psi_\pm=c^{(1)}_\pm \psi_\pm^{(1)} + c^{(2)}_\pm \psi_\pm^{(2)}: c^{(1)}_\pm, c^{(2)}_\pm \in\mathbb{C}\}$ are two dimensional; the deficiency indices are $n_+=n_-=n=2$. Consequently, the extended domain is parametrized by $2\times2$ unitary matrix maps 
\begin{equation}
\mathcal{U}:\Biggl(\begin{matrix} \psi_+^{(1)} \\ \psi_+^{(2)} \end{matrix}\Biggr)\mapsto\Biggl(\begin{matrix} u_{11} & u_{12} \cr u_{21} & u_{22} \end{matrix}\Biggr) \Biggl(\begin{matrix} \psi_-^{(1)} \\ \psi_-^{(2)} \end{matrix}\Biggr) \equiv u \Biggl(\begin{matrix} \psi_-^{(1)} \\ \psi_-^{(2)} \end{matrix}\Biggr).
\end{equation}
The domain of the self-adjoint extension is
\begin{eqnarray}\label{eq:dom}
\mathscr{D}(\mathcal{H_U})=\{ \psi = \phi+ \psi_++\mathcal{U} \psi_+: \phi_s(0)=\phi_s(L)=0,\ \psi_+\in\mathscr{K}_+\}.
\end{eqnarray} 

What boundary conditions characterize $\mathscr{D}(\mathcal{H_U})$? To find the answer, we note that for any $\chi,\psi\in\mathscr{D}(\mathcal{H_U})$, we must have
\begin{equation}\label{eq:HUsymm}
\langle\chi|\mathcal{H_U}\psi\rangle = \langle\mathcal{H_U}\chi|\psi\rangle.
\end{equation}
In particular, this must be true if we choose $\chi=\psi_+^{(1)}+\mathcal{U}\psi_+^{(1)}$ and $\chi=\psi_+^{(2)}+\mathcal{U}\psi_+^{(2)}$ separately. By partial integration, it is easy to see that
\begin{equation}
\langle \psi_\pm^{(a)} | \mathcal{H_U} \psi \rangle = \langle \mathcal{H_U} \psi_\pm^{(a)} | \psi \rangle + iv \left[{\psi_\pm^{(a)}(x)}^\dagger\sigma_y\psi(x)\right]_0^L, \quad a=1,2.
\end{equation}
The boundary terms are straightforward to calculate and yield
\begin{eqnarray}
i \left[{\psi_\pm^{(a)}(x)}^\dagger\sigma_y\psi(x)\right]_0^L &=& \exp\left (\frac{\zeta_a-1}2\frac{|\epsilon|}vL\right ) \left[\psi_\downarrow(L) + \zeta_ae^{\mp i \delta}\psi_\uparrow(L)\right] \nonumber \\ 
&& - \exp \left (-\frac{\zeta_a+1}2\frac{|\epsilon|}vL\right ) \left[\psi_\downarrow(0) + \zeta_a e^{\mp i \delta}\psi_\uparrow(0)\right] \nonumber\\
&\equiv& \pm\xi_\pm^{(a)},
\end{eqnarray}
with $\zeta_a = (-1)^a$. From Eq.~(\ref{eq:HUsymm}), collecting the boundary terms, we find the matrix equation $u\xi_+[\psi]=\xi_-[\psi]$, where 
$\xi_+[\psi]=( \xi_+^{(1)} \ \xi_+^{(2)})^{\sf T}$ and 
$\xi_-[\psi]=( \xi_-^{(1)} \ \xi_-^{(2)})^{\sf T}$. 
Since we know that the family of self-adjoint extensions is the same for all choices of $\eta$, we can rewrite this for a particular choice of $\eta$; namely we shall take the limiting case $\eta\to\infty$ (i.e., $|\epsilon|\to\infty$ and $e^{-i\delta}\to i$) to find
\begin{equation}\label{eq:U2BC}
u \Biggl(\begin{matrix} -\psi_\downarrow(0)+i\psi_\uparrow(0) \\ \psi_\downarrow(L)+i\psi_\uparrow(L) \end{matrix}\Biggr) = \Biggl(\begin{matrix} \psi_\downarrow(0)+i\psi_\uparrow(0) \\ -\psi_\downarrow(L)+i\psi_\uparrow(L) \end{matrix}\Biggr).
\end{equation}

The boundary condition can also be written in a form that appears independent of$~~\mathcal{U}$. This can be done by noting that the boundary conditions relate the two vectors $\xi_+$ and $\xi_-$ by a unitary matrix $u$. Equivalently, for two states $\psi,\phi\in\mathscr{D}(\mathcal{H_U})$ the inner product between the two vectors $\xi_\pm[\psi]$ and $\xi_\pm[\phi]$ defined separately for $\psi$ and $\phi$ must be preserved, i.e., $\xi_+[\psi]^\dagger\xi_+[\phi]=\xi_-[\psi]^\dagger\xi_-[\phi]$. After some algebra this yields
\begin{equation}\label{eq:jL0}
\psi_\uparrow^*(L)\phi_\downarrow(L) - \psi_\downarrow^*(L)\phi_\uparrow(L) = \psi_\uparrow^*(0)\phi_\downarrow(0) - \psi_\downarrow^*(0)\phi_\uparrow(0).
\end{equation}
This condition is quadratic in the wavefunctions and so is not immediately useful in solving the differential equations needed to find the spectrum. However, it is equivalent to the linear boundary conditions in Eq.~(\ref{eq:U2BC}). We note that it reduces to the boundary condition we found for the semi-infinite geometry if we let $L\to\infty$ and set $\psi_s(\infty)=0$. 

We next study the spectral symmetry as in the previous case. For the spin-flip operation, $\mathcal{S}$, we find $\xi_\pm[\mathcal{S}\psi] = \pm i\sigma_z\xi_\mp[\psi]$. Thus, using the fact that $u$ is unitary, the boundary condition is mapped to $u'\xi_+[\psi]=\xi_-[\psi]$ with $u'=-\sigma_zu^\dagger\sigma_z$. A boundary condition preserving the spectral symmetry is found when $u=u'$. This latter condition can be rewritten as $\mathcal{M}=\mathcal{M}^\dagger=\mathcal{M}^{-1}$ for $\mathcal{M}=i\sigma_z u$. That is, $\mathcal{M}$ is Hermitian and squares to~$\mathbf{1}$. There are two types of solutions to this equation: (1) $\mathcal{M}=\pm\mathbf{1}$, yielding $u=\mp i \sigma_z$; and (2) $\mathcal{M}=m_x\sigma_x+m_y\sigma_y+m_z\sigma_z\equiv \mathbf{m}\cdot \boldsymbol{\sigma}$, with a unit vector $\mathbf{m} =(m_x,m_y,m_z)\in S^2$ on the real two-sphere, yielding $u=-i \sigma_z (\mathbf{m}\cdot \boldsymbol{\sigma})$. It is not difficult to show that in the first case, the boundary conditions for the wavefunctions read
\begin{align}\label{eq:BC1}
\psi_\uparrow(0)=\pm\psi_\downarrow(0), \quad \psi_\uparrow(L)=\pm\psi_\downarrow(L).
\end{align}
As an example of the second case, we take $u=\mp i\mathbf{1}$. Then,
\begin{align}\label{eq:BC2}
\psi_\uparrow(0)=\pm\psi_\downarrow(0), \quad \psi_\uparrow(L)=\mp\psi_\downarrow(L).
\end{align}

The effect on the spectrum can be seen directly in terms of the number of zero-energy states in each case. With the boundary conditions in Eq.~(\ref{eq:BC1}) there is always \textit{one} exact zero-energy state, 
\begin{align}
\psi^{(+)}_{0}(x) = e^{-mx/v} \Biggl(\begin{matrix} 1 \\ 1 \end{matrix}\Biggr) \quad \text{or} \quad \psi^{(-)}_{0}(x) = e^{-m(L-x)/v} \Biggl(\begin{matrix} 1 \\ - 1 \end{matrix}\Biggr).
\end{align}
For the the boundary conditions in Eq.~(\ref{eq:BC2}) there are no exact zero-energy states for finite $L$. However, for large $L$ the conditions with the upper sign are satisfied by both $\psi_0^{(+)}$ \textit{and} $\psi_0^{(-)}$ to the order $e^{-mL/v}$, while for the lower sign the conditions are never satisfied independent of~$L$.  Thus, in this case, there are no zero-energy states for the lower sign and there are \textit{two} asymptotically exact zero-energy states for the upper sign.

Thus, we see that even after imposing the spectral symmetry given by $\mathcal{S}$, the bulk-boundary correspondence may still be incomplete. 
This can be understood in terms of the physical implications of the boundary conditions in Eqs.\ (\ref{eq:BC1}) and~(\ref{eq:BC2}). These conditions can be understood as modeling the interface of the wire with an environment. Note that the condition $\psi_\uparrow(0)=\psi_\downarrow(0)$ in the semi-infinite case allowed the existence of a zero-energy state localized at the edge, while the condition $\psi_\uparrow(0)=-\psi_\downarrow(0)$ did not. Therefore, the former condition models the interface of a normal insulator, the ``vacuum,'' with the \textit{left} edge of the wire that supports a topological bound state. If we had chosen the semi-infinite geometry $x<0$ with a \textit{right} edge, it is easy to see that the sign of the boundary condition modeling this situation would have been the opposite. This shows up in the finite geometry: now the boundary conditions in Eqs.\ (\ref{eq:BC1}) and~(\ref{eq:BC2}) at the two edges separately model an interface with vacuum with the $+$ sign on the left edge and the $-$ sign at the right edge.

\section{Self-adjoint extensions and conserved currents}\label{sec:curr}

As we have seen in several examples, such as in Eq.~(\ref{eq:jL0}), the conditions determining the domain of self-adjoint extensions of a Hermitian operator were written in a way that was independent of the specific choice of $\mathcal{U}$. In this section we show that this is quite a general result. Furthermore, we show that this can be directly interpreted as the conservation of a local current associated with the self-adjoint operator.

Given a self-adjoint operator $A$ and $\psi,\phi\in\mathscr{D}(A)$, we must have $\langle \psi | A\phi \rangle - \langle A\psi | \phi \rangle = 0$. This equation is the basis of the self-adjoint extensions we have so far discussed. For example, for the momentum operator $P=-id/dx$ on the Hilbert space $\mathscr{L}^2[0,L]$ we have
\begin{align}
0 &= \langle \psi | P\phi \rangle - \langle P\psi | \phi \rangle \nonumber \\
& = \int_0^{L}\left[\psi^*(x)\left(-i\frac{d}{dx}\phi(x)\right)-\left(-i\frac{d}{dx}\psi(x)\right)^*\phi(x)\right]dx \nonumber \\
&=-i\int_0^{L}\frac{d}{dx}\bigl(\psi^*(x)\phi(x)\bigr)\,dx.
\end{align}
As a result, we find $\psi^*(L)\phi(L)-\psi^*(0)\phi(0)=0$. The general solution independent of $\psi$ and $\phi$ is $\psi(L)=e^{i\beta}\psi(0)$ as we found under Eq.~(\ref{eq:Ptheta}). Note that for $\psi=\phi$ this condition reads
\begin{equation}
|\psi(L)|^2 = |\psi(0)|^2,
\end{equation}
which is simply the conservation of probability at the two boundaries. Since the momentum operator generates space translations, the probability density can be interpreted as the conserved current under space translations. Therefore, this condition simply states that the probability density must be conserved throughout the interval $[0,L]$.

Indeed, we show now that this way of specifying self-adjoint extensions of a Hermitian operator is quite general. For a self-adjoint operator $A$ one can always define a one-parameter family of unitary operators $U_A(\alpha) = e^{-i\alpha A}$, generated by $A$. This result is known as Stone's theorem.\cite{Reed-Simon1}  For example, the Hamiltonian generates the time-evolution operator with $\alpha$ being the time parameter. Then we can ``evolve'' any given state $\psi\in\mathscr{D}(A)$ as $\psi(\alpha)=U_A(\alpha)\psi$ with the initial condition $\psi(0)=\psi$. Since for two such evolved states, $\psi(\alpha)$ and $\phi(\alpha)$, the inner product $\langle \psi(\alpha) | \phi(\alpha) \rangle = \int \psi(x,\alpha)^*\phi(x,\alpha) \,dx$ is independent of $\alpha$, the density $\rho_A(x,\alpha) \equiv \psi(x,\alpha)^*\phi(x,\alpha) $ must have a \textit{local} conserved current, $j_A(x,\alpha)$, associated with it such that the following continuity equation is satisfied:
\begin{equation}
\frac{\partial}{\partial\alpha}\rho_A+\frac{\partial}{\partial x}j_A = 0.
\end{equation}
To see this, note that since $id\psi(\alpha)/d\alpha = A\psi(\alpha)$, we have
\begin{equation}
i \frac{\partial}{\partial\alpha}\rho_A(x,\alpha) = \psi^*(x,\alpha)(A\phi)(x,\alpha) - (A\psi)^*(x,\alpha)\phi(x,\alpha).
\end{equation}
Thus, the self-adjoint condition can be directly written as
\begin{align}
0 &= i \bigl[ \langle \psi | A\phi \rangle - \langle A\psi | \phi \rangle \bigr] \nonumber\\
&= - \int \frac{\partial}{\partial\alpha}\rho_A(x,\alpha) \,d x \nonumber\\
&= \int \frac{\partial}{\partial x} j_A(x,\alpha)\, dx.
\end{align}
Since this is satisfied for all $\alpha$, and in particular $\alpha=0$, we find that the conserved current
\begin{equation}
j_A(x) = i \int^x \bigl[ \psi^*(x')(A\phi)(x') - (A\psi)^*(x')\phi(x') \bigr] \,dx'
\end{equation}
characterizes the domain of the self-adjoint operator $A$.

For the continuum Hamiltonian $\mathcal{H}$, Eq.~(\ref{eq:H}), one finds the local current
\begin{equation}\label{eq:current}
j_{\mathcal{H}}(x) = -v \psi(x)^\dagger\sigma_y\phi(x) = iv\bigl[ \psi_\uparrow^*(x)\phi_\downarrow(x) - \psi_\downarrow^*(x)\phi_\uparrow(x) \bigr].
\end{equation}
So, in the semi-infinite geometry, noting that $j_{\mathcal{H}}(\infty)=0$, we find $j_{\mathcal{H}}(0)=0$. This reads $\psi^*_\uparrow(0)\phi_\downarrow(0)=\psi^*_\downarrow(0)\phi_\uparrow(0)$, which indeed yields Eq.~(\ref{eq:BCsemi}) independent of the states.
In the finite geometry $x\in[0,L]$ we have $j_{\mathcal{H}}(L) = j_{\mathcal{H}}(0)$, which is precisely the condition we found in Eq.~(\ref{eq:jL0}).
In order to obtain the boundary conditions from the current condition, we must revert the steps that took us from Eq.~(\ref{eq:U2BC}) to Eq.~(\ref{eq:jL0}). To do so, we use the identity $j_{\mathcal{H}}=\frac12v\left[(-\psi_\downarrow+i\psi_\uparrow)^*(-\phi_\downarrow+i\phi_\uparrow)-(\psi_\downarrow+i\psi_\uparrow)^*(\phi_\downarrow+i\phi_\uparrow)\right]$ to write $j_{\mathcal{H}}(L) = j_{\mathcal{H}}(0)$ as
\begin{equation}\label{eq:jHinner}
\Biggl(\begin{matrix}
-\psi_\downarrow(0)+i\psi_\uparrow(0) \\
\psi_\downarrow(L)+i\psi_\uparrow(L)
\end{matrix}\Biggr)^\dagger
\Biggl(\begin{matrix}
-\phi_\downarrow(0)+i\phi_\uparrow(0)\\
\phi_\downarrow(L)+i\phi_\uparrow(L)
\end{matrix}\Biggr)
=
\Biggl(\begin{matrix}
\psi_\downarrow(0)+i\psi_\uparrow(0) \\
-\psi_\downarrow(L)+i\psi_\uparrow(L)
\end{matrix}\Biggr)^\dagger
\Biggl(\begin{matrix}
\phi_\downarrow(0)+i\phi_\uparrow(0)\\
-\phi_\downarrow(L)+i\phi_\uparrow(L)
\end{matrix}\Biggr).
\end{equation}
This means that the boundary conditions must map
\begin{equation}
\Biggl(\begin{matrix}
-\psi_\downarrow(0)+i\psi_\uparrow(0)\\
\psi_\downarrow(L)+i\psi_\uparrow(L)
\end{matrix}\Biggr)
\mapsto
\Biggl(\begin{matrix}
\psi_\downarrow(0)+i\psi_\uparrow(0)\\
-\psi_\downarrow(L)+i\psi_\uparrow(L)
\end{matrix}\Biggr),
\end{equation}
such that the inner product in Eq.~(\ref{eq:jHinner}) is preserved independently of the state $\psi$. This is satisfied if and only if the states satisfy the boundary condition in Eq.~(\ref{eq:U2BC}). This method can be used to obtain the boundary conditions from the conserved current more generally; see the discussion in Appendix~\ref{app:jBC}.

\section{Conclusion}\label{sec:conc}

As elaborated in this paper, the study of the continuum limit of a lattice model is a powerful method for extracting the universal, long-distance physics of the system. Different lattice models can share the same continuum, or field-theory, description and, therefore, the same universal behavior. 
In the context of topological phases, it is instructive to compare our results to those for the Kitaev model for a one-dimensional $p$-wave superconductor.~\cite{Kitaev,EllFra15a} This is a model proposed for the realization of exotic topological bound states, known as Majorana fermions, with special non-Abelian braiding properties potentially useful for quantum computation. It has recently received great attention as experimental efforts to realize the model in nanowires have produced encouraging results.~\cite{EllFra15a} While the physics of this system is quite different from the one we studied in this paper (e.g., there is no superconductivity in our system), the continuum limits of the two systems are equivalent. This means they have the same topological phase diagram. In particular, the self-adjoint extensions in the continuum and the topological bound states are in one-to-one correspondence to each other. What differs from one continuum model to the other is the specific structure of the quantum fields and, most importantly, the physical properties of the different phases. In fact, it turns out that these two models are not only equivalent in the continuum limit but that their lattice versions are unitarily equivalent (see Appendix~\ref{app:FMK} for a proof). So, our lattice and field theory analyses are directly applicable to the study of topological phases of the Kitaev model. 

Topological bound states, such as Majorana fermions, are interesting in part for their usefulness in device applications. These applications are often based on dynamical features of the bound states as they are moved around or otherwise manipulated. In such manipulations, the Hamiltonian of the system is changed in time by varying the external parameters of the system. For example, by changing the Zeeman energy $\mu$ one can manipulate the position of the bound states. It is worth emphasizing that, when analyzing such schemes in the continuum, it is important to work with a self-adjoint extension of the Hamiltonian since only such an operator can define the time evolution of the system properly. 

In summary, through a simple model we have illustrated the importance of self-adjoint extensions in the continuum, or field-theory, description of topological phases of quantum systems with boundaries. In particular, we clarified the physical interpretation of the extended operators in terms of a conserved local current. These extensions correspond to different physical situations with physically distinct environments outside the system, or equivalently, experimental conditions. We showed that the distinction can persist even after imposing internal symmetries that restrict the choice of the extension. Thus, the notion of bulk-boundary correspondence in a topological phase cannot in general be defined independently from the choice of the extension and the corresponding boundary conditions at the edges of the system.

\begin{acknowledgments}
This work was supported by the NSF through Grant No. DMR-1350663 (B.S. and M.T.A.), by the BSF Grant No. 2014345 (B.S.), and by the College of Arts and Sciences at Indiana University.
\end{acknowledgments}

\appendix

\section{Continuum ladder operators}\label{app:ladderCL}

Here we show that the operator $\hat\Psi_E$ defined in Eq.~(\ref{eq:psiE}) is indeed a ladder operator satisfying the commutation relation $[\hat H_c,\hat\Psi^\dagger_E]=E\hat\Psi^\dagger_E$. Using the algebraic relation $[\hat{A}\hat{B},\hat{C}] = \hat{A}(\hat{B}\hat{C} + \hat{C}\hat{B}) - (\hat{A}\hat{C}+\hat{C}\hat{A})\hat{B}$ and the anticommutation relations~(\ref{eq:commCL}), we have
\begin{align}
[\hat H_c,\hat\Psi^\dagger_E] 
&= \sum_{s_1 s_2 s'}\int \left[\hat\Psi^\dagger_{s_1}(x)\mathcal{H}_{s_1s_2}\hat\Psi_{s_2}^{\vphantom{\dagger}}(x),\phi_{s' E}(x')\hat\Psi^\dagger_{s'}(x')\right] dx\,dx' \nonumber\\
&= \sum_{s_1 s_2 s'}\int \mathcal{H}_{s_1s_2}\phi_{s' E}(x')\hat\Psi^\dagger_{s_1}(x) \left[\hat\Psi_{s_2}(x)^{\vphantom{\dagger}}\hat\Psi^\dagger_{s'}(x')+\hat\Psi^\dagger_{s'}(x')\hat\Psi_{s_2}^{\vphantom{\dagger}}(x)\right] dx\,dx' \nonumber\\
&= \sum_{s_1 s_2 s'}\int \mathcal{H}_{s_1s_2}\phi_{s' E}(x')\hat\Psi^\dagger_{s_1}(x) \delta_{s_2 s'}\delta(x-x')\, dx\, dx'\nonumber\\
&= \sum_{s_1 s'}\int \mathcal{H}_{s_1s'}\phi_{s' E}(x) \hat\Psi^\dagger_{s_1}(x)\,dx\nonumber \\
&= \sum_{s_1}\int E\phi_{s_1 E}(x) \hat\Psi^\dagger_{s_1}(x)\,dx \nonumber\\
&= E \hat\Psi^\dagger_E,
\end{align}
where, in the penultimate line, we also used Eq.~(\ref{eq:1stH}).

\section{Dense and closed operators}\label{app:dense}

In this appendix we give definitions for densely defined and closed operators. We illustrate these concepts with the momentum operator $P$.

\begin{quote}
\textbf{Definition 1.}
A subset $\mathscr{S}\subset\mathscr{H}$ is \textit{dense} if for every $\psi\in\mathscr{H}$, there is a sequence $\psi_n\in\mathscr{S}$ that converges to $\psi$ in norm, written simply $\psi_n\to\psi$, i.e., $\lim_{n\to\infty}\| \psi_n-\psi \| = 0$.
\end{quote}

\begin{quote}
\textbf{Definition 2.}
An operator $A:\mathscr{D}(A)\to\mathscr{H}$ is a \textit{densely defined operator} if $\mathscr{D}(A)$ is dense in $\mathscr{H}$.
\end{quote}

Equivalently, an operator is densely defined if there is an orthonormal basis $\phi_j\in\mathscr{D}(A)$ for the Hilbert space such that any state $\psi\in\mathscr{H}$ can be written uniquely as a superposition of $\phi_j$, that is, $\psi=\sum_j a_j\phi_j$, with $a_j=\langle\phi_j|\psi\rangle$. Here, the sum $\sum_j$ is formally understood as a sum over discrete values of $j$ and an integral over continuous values of $j$. 

An operator $A$ must be densely defined in order to uniquely define its adjoint, $A^\dagger$, with the relation $\langle \phi | A^\dagger\psi \rangle=\langle A \phi | \psi \rangle$; when this property holds for all $\phi_j$, $A^\dagger\psi$ is uniquely written as $A^\dagger\psi=\sum_j \langle \phi_j | A^\dagger\psi \rangle \phi_j = \sum_j\langle A \phi_j | \psi \rangle \phi_j$.

\begin{quote}
\textbf{Definition 3.}
The operator $A$ is \textit{closed} if for any sequence $\psi_n\in\mathscr{D}(A)$ for which
$\psi_n \to \psi$ and $A\psi_n \to \phi$, we have $\psi\in\mathscr{D}(A)$ and $\phi=A\psi$.
\end{quote}


Let us show that the momentum operator $P=-id/dx$, with
\begin{equation}
\mathscr{D}(P) = \{ \psi\in\mathscr{L}^2[0,L]:\psi \text{ is absolutely continuous}, \psi(0)=\psi(L)=0 \},
\end{equation}
is densely defined and closed. First note that there is an orthonormal basis $\phi_j\in\mathscr{D}(P)$, $j\in\mathbb{N}$, given by $\phi_j(x)=\sqrt{2/L}\sin(j \pi x/L)$. For any $\psi\in\mathscr{H}$, the sequence of partial Fourier series
$\psi_n$ defined by
\begin{equation}
\psi_n(x)=\sum_{j=1}^{n}{a_j\phi_j(x)}\quad \text{with}\ a_j=\int_0^{L}\phi_j^*(x)\psi(x) \,dx
\end{equation}
converges to $\psi$ in norm. Thus, $\mathscr{D}(P)$ is dense in $\mathscr{L}^2[0,L]$. 

To show that $P$ is closed, we assume a sequence $\chi_n\in\mathscr{D}(P)$ is given so that $\chi_n\to\psi$ and $P\chi_n\to\phi$. Since $\mathscr{D}(P)$ is dense, we may write $\psi=\sum_ja_{j}\phi_j$ and define the partial-sum sequence $\psi_n\equiv\sum_j^na_{j}\phi_j\in\mathscr{D}(P)$ that also converges to $\psi$, $\psi_n\to\psi$. By continuity, $P\psi_n\to\phi$, too. Since $\lim_{n\to\infty}P\psi_n=-i\lim_{n\to\infty}\sum_j^na_{j}\,d\phi_j/dx$ exists, $\lim_{n\to\infty}\psi_n=\psi$ is absolutely continuous and vanishes at the boundaries, $x=0,L$;~\cite{Rudin} therefore, $\psi_n\to\psi\in\mathscr{D}(P)$. Finally, we show that $P\psi_n\to P\psi$ and, thus, $\phi=P\psi$:
\begin{align}
-i\frac{d}{d x}\psi_n(x) &=-i\frac{\pi}{L}\sqrt{\frac{2}{L}}\sum_{j=1}^n j a_{j}\cos{\frac{j\pi x}{L}} \nonumber \\
&= -i\frac{2}{L}\frac{\pi}{L}\sum_{j=1}^n \left[\int_0^{L} j\sin{\frac{j\pi x'}{L}}\psi(x')\,dx' \right]\cos{\frac{j\pi x}{L}} \nonumber\\
&= +i\frac{2}{L}\sum_{j=1}^n \left[\int_0^{L} \left(\frac{d}{d x'}\cos{\frac{j\pi x'}{L}}\right)\psi(x')\,dx'\right]\cos{\frac{j\pi x}{L}} \nonumber\\
&= -i\int_0^{L}\left[\frac{2}{L}\sum_{j=1}^n \cos{\frac{j\pi x'}{L}}\cos{\frac{j\pi x}{L}}\right]\frac{d}{d x'}\psi(x')\,dx';
\end{align}
since
\begin{equation}
\lim_{n\to\infty}\frac{2}{L}\sum_{j=1}^n \cos{\frac{j\pi x'}{L}}\cos{\frac{j\pi x}{L}}=\delta(x-x')-\frac{1}{L},
\end{equation}
we conclude that 
\begin{equation}\label{eq:Pf}
\lim_{n\to\infty}\| P\psi_n - P\psi\|=0.
\end{equation}

\section{Boundary conditions from conserved current condition}\label{app:jBC}

In order to find the boundary conditions from the conserved current of Sec.~\ref{sec:curr}, we note that the current $j_A$ is a local, sesquilinear, Hermitian form of the two states $\psi,\phi$, such that $j_A(\psi,\phi)=j_A(\phi,\psi)^*$. A general expression for such a form is $j_A(x)=\nu[\psi(x)]^\dagger\, J_A \, \nu[\phi(x)]$, where $\nu[\psi(x)]$ is an $n$-dimensional vector of linear combinations of the elements of $\psi(x)$ and its derivatives, and $J_A^\dagger=J_A^{\vphantom{\dagger}}$ is a Hermitian matrix. For example, for the momentum operator $P=-id/dx$, we have $\nu[\psi(x)]=\psi(x)$, $n=1$, $J_P=1$.

\begin{quote}
\textbf{Exercise 2.}
For the operator $\Delta=-d^2/dx^2$, show that 
$\nu[\psi]=(\psi \ \, d\psi/dx)^{\sf T}$, 
$J_\Delta=\sigma_y$, and $n=2$.
\end{quote}

We can diagonalize $J_A=T^\dagger D T$, with $T$ unitary and $D$ diagonal with real elements. The unitary matrix $T$ can be always chosen such that the diagonal elements of $D$ are sorted in decreasing order. Then $D = (d_+ \oplus d_-)^2$, with $\oplus$ the direct sum, such that $d_\pm = \pm d_\pm^\dagger$ are Hermitian and anti-Hermitian, respectively, and diagonal; the diagonal elements of $d_\pm$ are simply the square roots of, respectively, the positive and negative elements of~$D$.

\begin{quote}
\textbf{Exercise 3.}
For the continuum Hamiltonian, $\mathcal{H}=iv\sigma_y d/dx+m\sigma_z$, show that $\nu[\psi]=\psi$, $J_{\mathcal{H}}=-v\sigma_y$, and $n=2$. Also, show that $T=e^{i\pi\sigma_x/4}=(\mathbf{1}+i\sigma_x)/\sqrt2$, 
\begin{equation}
D=v\sigma_z=\Biggl(\begin{matrix} \sqrt v & 0 \\ 0 & -i\sqrt v \end{matrix}\Biggr)^2,
\end{equation}
and, thus, $d_+=\sqrt{v}$, $d_-=-i\sqrt{v}$ (up to signs). 
\end{quote}

Now, writing $J_A=T^\dagger(d_+\oplus d_-)^2T=\big[(d_+\oplus d_-)^\dagger T\big]^\dagger\big[(d_+\oplus d_-)T\big]$, we can see that
\begin{align}
j_A 
&= \big((d_+ \oplus - d_-)T\,\nu[\psi]\big)^\dagger \big((d_+ \oplus d_-)T\,\nu[\phi]\big)\nonumber \\
&= \nu_+[\psi]^\dagger\nu_+[\phi]^{\vphantom{\dagger}} - \nu_-[\psi]^\dagger\nu_-[\phi]^{\vphantom{\dagger}},
\end{align}
where we have defined
\begin{equation}
(d_+ \oplus d_-)T\,\nu[\psi] =: \nu_+[\psi]\oplus\nu_-[\psi].
\end{equation}

We will now consider two different geometries. First, consider a semi-infinite geometry $x\in[0,\infty)$.  Then the current condition $j_A(0)=0$ reads
\begin{equation}
\nu_+[\psi(0)]^\dagger\nu_+[\phi(0)] = \nu_-[\psi(0)]^\dagger\nu_-[\phi(0)].
\end{equation}
If, and only if, $d_+$ and $d_-$ (and, thus, $\nu_+$ and $\nu_-$) have the same dimension, $n/2$, this condition has a solution in terms of a unitary $\frac n2\times\frac n2$ matrix $\bar{u}$ that maps $\nu_+\mapsto\nu_-$, i.e.,
\begin{equation}\label{eq:jBCsemi}
\bar u\,\nu_+[\psi(0)]=\nu_-[\psi(0)].
\end{equation}
Otherwise, $A$ will have no self-adjoint extensions. For example, for the momentum operator $P=-id/dx$, we have $\nu_+[\psi(0)]=\psi(0)$ and $\nu_-$ does not exist. Therefore, $P$ cannot have a self-adjoint extension on the semi-infinite line. For the continuum Hamiltonian, $\mathcal{H}$, we have $\nu_\pm[\psi(0)]=\sqrt{v/2}\left[\psi_\uparrow(0)\pm i\psi_\downarrow(0)\right]$. Therefore, the self-adjoint extensions of $\mathcal{H}$ are characterized by a U(1) phase $\bar u = e^{i\theta}$; after simple algebra, Eq.~(\ref{eq:jBCsemi}) reads $\psi_\uparrow(0)=\cot(\theta/2)\psi_\downarrow(0)$, which is the same as Eq.~(\ref{eq:BCsemi}) with $\delta=\pi/2$.

Now consider a finite geometry $x\in[0,L]$. The current condition, $j_A(L)=j_A(0)$, yields
\begin{align}
\nu_+[\psi(0)]^\dagger\nu_+[\phi(0)] + \nu_-[\psi(L)]^\dagger\nu_-[\phi(L)] 
&= 
\nu_-[\psi(0)]^\dagger\nu_-[\phi(0)] + \nu_+[\psi(L)]^\dagger\nu_+[\phi(L)] \\
\Rightarrow \quad
\varsigma_+[\psi]^\dagger\varsigma_+[\phi] &= \varsigma_-[\psi]^\dagger\varsigma_-[\phi],
\end{align}
where
\begin{equation}
\varsigma_\pm[\psi] \equiv \nu_\pm[\psi(0)] \oplus \nu_\mp[\psi(L)].
\end{equation}
Thus, the state-independent boundary conditions are given by a norm-preserving, $n\times n$ unitary matrix $u$ that maps $\varsigma_+ \mapsto \varsigma_-$, i.e.,
\begin{equation}\label{eq:jBCfinite}
u\,\varsigma_+[\psi] = \varsigma_- [\psi].
\end{equation}
In the case of the momentum operator $P=-id/dx$ we see immediately that now $\varsigma_+[\psi]=\psi(0)$ and $\varsigma_-[\psi]=\psi(L)$. Thus, the self-adjoint extensions of $P$ are characterized by a U(1) phase as $\psi(L)=e^{i\beta}\psi(0)$. In the case of the continuum formulation of the topological wire, we find 
\begin{equation}
\varsigma_+=-i{\sqrt \frac v2}\Biggl(\begin{matrix}  -\psi_\downarrow(0) + i \psi_\uparrow(0) \\ \psi_\downarrow(L) + i \psi_\uparrow(L)  \end{matrix}\Biggr),
\quad
\varsigma_-= -i{\sqrt \frac v2} \Biggl(\begin{matrix}  \psi_\downarrow(0) + i \psi_\uparrow(0) \\ - \psi_\downarrow(L) + i \psi_\uparrow(L)  \end{matrix}\Biggr).
\end{equation}
Thus, from Eq.~(\ref{eq:jBCfinite}) we find the boundary conditions in Eq.~(\ref{eq:U2BC}).

\section{Unitary equivalence between a ferromagnetic insulator and a $p$-wave superconductor}\label{app:FMK}

Here we prove that the ferromagnetic insulator model in Eq.~(\ref{eq:Hlatt}) is unitarily equivalent to a $p$-wave superconductor. More precisely, the superconducting model represents two uncoupled 
Kitaev chains~\cite{Kitaev} that share identical energy spectra. 

Define the local canonical map
\begin{eqnarray}
c^\dagger_{\uparrow r}=\frac{f^\dagger_r + i g^\dagger_r}{\sqrt{2}} , \quad
c^\dagger_{\downarrow r}=\frac{f^{\;}_r + i g^{\;}_r}{\sqrt{2}} , 
\end{eqnarray}
in terms of two new sets of fermion creation and annihilation operators, 
$f^\dagger_r$ and $f^{\;}_r$, and $g^\dagger_r$ and $g^{\;}_r$. It is straightforward to show that the $f$~operators commute with the $g$~operators and each pair separately satisfies fermionic anti-commutation relations $\{f^{\;}_r,f^\dagger_{r'}\} = \{g^{\;}_r,g^\dagger_{r'}\}  = \delta_{rr'}$, etc. Using these relation and after some algebraic manipulations, 
the Hamiltonian $\hat H$ of Eq.~(\ref{eq:Hlatt}) becomes
\begin{equation}
\hat{H}+\mu N=\sum_{r=1}^{N}\left[
\frac{\mu}{2}  (f_{r}^{\dagger}f_{r}^{\vphantom{\dagger}} + g_{r}^{\dagger}g_{r}^{\vphantom{\dagger}} )
+ w ( f_{r}^{\dagger}f_{r+1}^{\vphantom{\dagger}}+ g_{r}^{\dagger}g_{r+1}^{\vphantom{\dagger}})
+ \frac\lambda2 ( f_{r}^{\dagger}f_{r+1}^\dagger + g_{r}^{\dagger}g_{r+1}^\dagger) \right ]
+ \text{h.c.},
\end{equation}
where, as before, the sum is understood not to contain any terms with the index $N+1$. Thus, the Hamiltonian is mapped to two copies of the Kitaev model for $f$ and $g$ separately.


\begin{thebibliography}{10}
\bibitem{Transpose}
The transpose operation ${\sf T}$ is defined as
$$
( \phi_{1}(x) \ \phi_{2}(x) )^{\sf T} = 
\Biggl(\begin{matrix} \phi_{1}(x) \\ \phi_{2}(x) \end{matrix}\Biggr) .
$$
\bibitem{Note1-5}
An absolutely continuous function $\phi$ on an interval $I\subset\mathbb{R}$ is defined as one for which there exists an integrable function $\phi'$ that satisfies  $\int_a^b\phi'(x) \, d x=\phi(b)-\phi(a)$ for any $a,b\in I$ (see, for example, p. 22 in Ref.~\onlinecite{GitTuyVor12a}). Then, according to the fundamental theorem of calculus, $\phi'=d\phi/d x$ almost everywhere in $I$.
\end{thebibliography}
\end{document}